\documentclass[aps,pra,twocolumn,nofootinbib]{revtex4-2}
\usepackage[colorlinks]{hyperref}
\usepackage[utf8]{inputenc}
\usepackage{graphicx}
\usepackage{multirow}
\usepackage{amssymb}
\usepackage{amsmath}
\usepackage{gensymb}

\begin{document}

\title{
Radiation from cold molecular clouds and Sun chromosphere \\ produced by anti-quark nugget dark matter
}

\author{V.~V.~Flambaum$^{1,2,3}$}
\author{I.~B.~Samsonov$^1$}
\affiliation{$^1$School of Physics, University of New South Wales, Sydney 2052, Australia}
\email{v.flambaum@unsw.edu.au, igor.samsonov@unsw.edu.au}
\affiliation{$^2$Helmholtz Institute Mainz, GSI Helmholtzzentrum für Schwerionenforschung, 55099 Mainz, Germany}
\affiliation{$^3$Johannes Gutenberg University Mainz, 55099 Mainz, Germany}

\begin{abstract}
We study astrophysical implications of the quark nugget model of dark matter and propose observational techniques for detecting anti-Quark Nuggets (anti-QNs) with modern telescopes. Anti-QNs are compact composite objects of antiquark matter with a typical radius $R\sim 10^{-5}$ cm and density exceeding that of nuclear matter. Atoms and molecules of interstellar medium collide with anti-quark nuggets and annihilate. We estimate thermal radiation from anti-QNs in cold molecular clouds in our galaxy and show that this radiation appears sufficiently strong to be observed in infrared and visible spectra. Proton annihilation on anti-QNs produces $\gamma$-photons with energies in the range 100-400 MeV which may be detected by telescopes such as Fermi-LAT. We have found that anti-QN radiation inside the solar corona is too weak to produce a significant plasma heating or any other observable effects, while the radiation of $\gamma$-photons from the chromosphere may be observable. We also address the problem of survival of anti-quark nuggets in the early universe. 
\end{abstract}

\maketitle

\section{Introduction}

The nature and fundamental properties of dark matter particles remain unknown despite the long history of these problems. Current experiments and observations do not allow us to give a preference to either of existing models of dark matter. It is an important task to study observational implications of each particular model and compare them with the results of experiments to select more probable candidates for dark matter and push aside less realistic ones.

In this paper, we study implications of the Quark Nugget (QN) model of dark matter particles basing upon the properties of this model developed in Refs.~\cite{FS21,FS2}. This model belongs to a class of dark matter models in which dark matter particles are represented by compact composite objects consisting of the standard model particles such as quarks and leptons. The well-known example of such models is the so-called strangelet \cite{Witten84}, a droplet of quark matter stabilized by s quark. 

Recently, there has been a surge of interest to this class of models owing to a series of seminal papers by A.~Zhitnitsky and his collaborators \cite{Zhitnitsky2002,Zhitnitsky2006,Electrosphere2008,WMAPhaze,corona1,corona2,corona3,FRB,FlambaumZhitnitsky1,FlambaumZhitnitsky2,BFZ} (see also \cite{Zhitnitsky-review} for a review), where the so-called Axion-Quark-Nugget (AQN) model was proposed and developed. This model has a few important features which make it very attractive for applications. The main new ingredient in this model is the (hypotetical) axion-pion domain wall, which keeps the quark matter under high pressure and prevents it from decays. As is argued in Refs.~\cite{Zhitnitsky2002,AQNformation,Survival}, this domain wall played also crucial role in formation of AQNs in the early universe. 

Another very important prediction of the AQN model is that dark matter particles may be represented by both quark nuggets built of quarks and leptons and anti-quark nuggets, which consist of antiquarks and antileptons. Assuming the asymmetry in production of these particles in the early universe, it is possible to explain matter-antimatter asymmetry at present in a very elegant way \cite{Zhitnitsky2006}: If anti-QNs are 1.5 times more abundant than QNs, then all antimatter is hidden inside anti-QNs, and the ratio of dark matter to visible matter mass contribution in the universe appears close to the observed one 5:1. Thus, in total, the baryon symmetry is preserved in the universe at all times.

Prediction of nuggets of antimatter (anti-QNs) makes the quark nugget model of dark matter especially attractive. In contrast with QNs, the anti-QNs strongly interact with visible matter and produce specific pattern of radiation \cite{Electrosphere2008,WMAPhaze,FS21,FS2}. Although the anti-QN annihilation events with visible matter are rare because of a 
small number density of DM particles, it could be detected with modern telescopes. In this paper, we will focus only on radiation from anti-QNs, as they serve as a unique tool for justification of this model.

The interest to the AQN model is also motivated by its success in resolving many unexplained phenomena in astrophysics and particle physics, such as primordial lithium abundance problem \cite{FlambaumZhitnitsky1}, pattern of radiation from our galaxy center \cite{Electrosphere2008,WMAPhaze}, solar corona temperature mystery \cite{corona1,corona2,corona3}, fast radio bursts \cite{FRB}, and others, see, e.g., Ref.~\cite{Zhitnitsky-review} for a review. AQN model proved also useful in explaining some mysterious phenomena in the Earth atmosphere and underground \cite{FlambaumZhitnitsky2,BFZ}.

In sections \ref{sec-radiation} and \ref{sec-nonthermal}, we study the properties of radiation from anti-QNs interacting with the gas in cold molecular clouds in our galaxy. As a typical molecular cloud we consider the Taurus molecular cloud, since it is one of the nearest ones, and its structure is well studied. We estimate the spectrum and radiation power from anti-QNs in this cloud and compare them with observations. 

In sections \ref{sec-corona} and \ref{sec-chromosphere}, we estimate the radiation from anti-QNs in the solar atmosphere. First, we reconsider the results of the works \cite{corona1,corona2,corona3} and show that anti-QN annihilation cannot produce significant effects on the solar corona heating. Then, we show that $\gamma$-ray radiation from anti-QN annihilation in chromosphere could be observed by modern satellites and observatories. 

Finally, in Sect.~\ref{sec-survival} we revisit the condition of survival of anti-QNs in hot plasma of early universe. We fix some of the factors omitted in earlier estimates \cite{Survival} and present a stronger constraint on the baryon charge required for survival of anti-QNs to the present day. We speculate also on the role of axion-pion domain wall in suppressing the annihilation cross section and allowing anti-QNs to survive after the QCD phase transition epoch. Note that to support equal number of quarks and anti-quarks in the universe, the total anti-barionic charge  of anti-QN should be  three times  bigger than the total barionic charge  of nucleons \cite{Zhitnitsky2002}. Therefore, two thirds of  anti-QNs survive even if all nucleons are annihilated. Hence, it may be more appropriate to talk about survival of ordinary matter rather than survival of anti-QNs.

In this work, we use natural units in which $\hbar =1$, $c=1$.


\section{Thermal radiation from anti-quark nuggets in giant molecular clouds}
\label{sec-radiation}

Anti-QNs emit potentially observable diffuse thermal radiation when they interact with the interstellar medium in molecular clouds in our galaxy. In this section, we start with a short summary about properties of thermal radiation of anti-QNs studied in Ref.~\cite{FS2}. Then, we calculate the corresponding radiation from the Taurus molecular cloud, which is one of the nearest molecular clouds to the Earth.

\subsection{Radiation from one quark nugget}

An anti-QN consists of an antiquark core and a positron cloud which compensates the electric charge of the core \cite{Zhitnitsky2002,Zhitnitsky2006,Electrosphere2008}. The antiquark core is supposed to possess a large baryon charge $B$ and the density exceeding that of the nuclear matter. We will assume that the radius of the quark nugget is 
\begin{equation}
    R= B^{1/3} {\rm fm}.
\end{equation}
The baryon charge number $B$ is a free parameter in this model which is constrained by $10^{24}\lesssim B\lesssim 10^{28}$, see, e.g., Ref.~\cite{Zhitnitsky-review} for a review.

The antiquark core may have an electric charge $|Q|\sim 10^{21}$. This charge should be compensated by the positron cloud around the antiquark core. The distribution of the electric charge in this cloud was studied in Refs.~\cite{WMAPhaze,Electrosphere2008,FS21,FS2}. At non-zero temperature, the thermal fluctuations of density in this cloud produce the thermal radiation from quark nuggets. The spectrum of this radiation was studied in Ref.~\cite{FS2}.

An anti-QN may be considered as a small particle characterized by the dielectric constant $\varepsilon(\omega)$
\begin{equation}
    \varepsilon(\omega) = 1-\frac{\omega_p^2}{\omega^2 + i\gamma \omega}\,,
    \label{epsilon}
\end{equation}
where $\omega_p\simeq2$ MeV is the plasma frequency and $\gamma\simeq0.5$ keV is the damping constant \cite{FS2}. 
The thermal radiation from quark nuggets is produced by fluctuations of density in the positron cloud around the antiquark core. The radiation power from unit surface area of QN per unit frequency interval is given by 
\begin{equation}
    P(\omega,T) = \pi E(\omega) I_0(\omega,T)\,,
    \label{P}
\end{equation}
where $I_0$ is the Plank function 
\begin{equation}
    I_0(\omega,T) = \frac{\hbar\omega^3}{4\pi^3 c^2}
    \frac1{\exp(\hbar\omega/(k_B T))-1}\,,
    \label{I0}
\end{equation}
and $E(\omega)$ is the QN emissivity function. For a wide range of frequencies, this function was studied analytically and numerically in Ref.~\cite{FS2}. Although, in general, this function is complicated, for low frequencies, $(\omega R/c)^2\ll 1$, it may be approximated by a simple expression:
\begin{equation}
    E(\omega) \approx  6 \,  {\rm Re}[(\varepsilon(\omega))^{-1/2}] = 6 {\rm Re} \left(\frac{\omega^2 + i \gamma \omega}{\omega^2 -\omega_p^2 + i \gamma \omega}\right)^{1/2}.
    \label{Egeneral}
\end{equation}
In the range of frequencies $\omega \ll \gamma \ll \omega_p$  
this expression may be further simplified
\begin{equation}
    E(\omega) \approx   3  \frac{\sqrt{2\omega \gamma}}{\omega_p}\,.
    \label{Eapprox}
\end{equation}

\begin{figure}[htb]
    \centering
    \includegraphics[width=8cm]{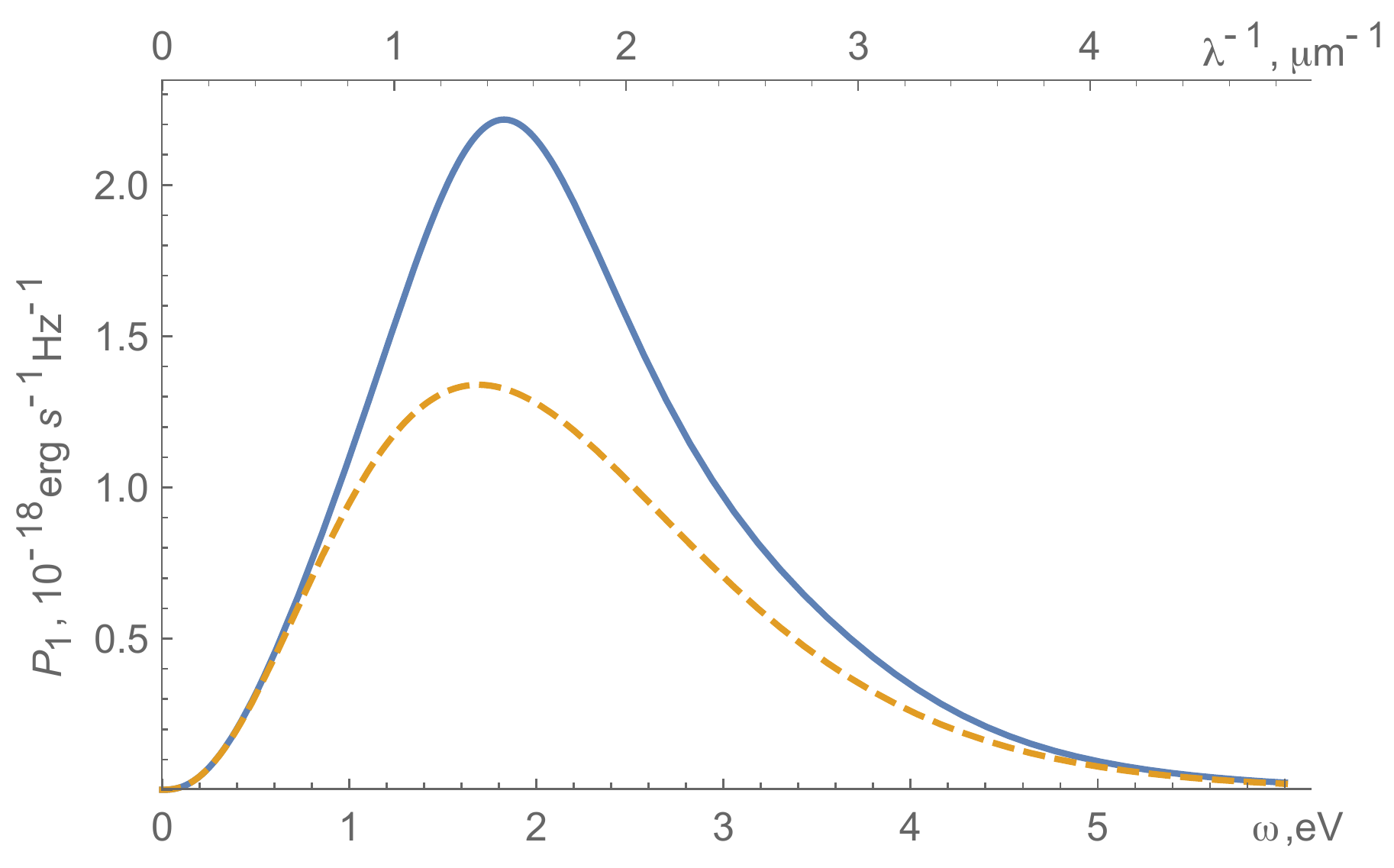}
    \caption{Spectrum of radiation of quark nuggets at temperature $T=0.5$ eV. Dashed curve represents the same spectrum calculated with the use of the approximate formula (\ref{Eapprox}).}
    \label{fig-radiation1}
\end{figure}

The thermal radiation spectrum of one (anti)quark nugget is found by multiplying Eq.~(\ref{P}) by the quark nugget area $4\pi R^2$,
\begin{equation}
    P_{1}(\omega,T) = 4\pi R^2 P(\omega,T)
    =4\pi^2 {\rm fm}^2 B^{2/3} E(\omega) I_0(\omega,T)\,.
    \label{P1}
\end{equation}
Here we expressed the QN radius via the baryon number, $R =  B^{1/3}{\rm fm}$. 
In Fig.~\ref{fig-radiation1}, we show a typical profile of this function for a particular value of QN effective temperature $T=0.5$ eV and baryon charge number $B=10^{24}$. At this temperature, the maximum of QN radiation is near the frequency $\omega = 2$ eV ($\lambda=620$ nm). Note that the approximate formula (\ref{Eapprox}) for the QN thermal emissivity gives the underestimated result for the radiation power by about a factor of 1.5.

\subsection{Radiation from Taurus molecular cloud}

Taurus molecular cloud is one of the best studied ones because of its proximity to  Earth, $L\simeq 140$ pc, and a large visible area covering more than 200 pc$^2$. The distribution of gas in this cloud was studied in Ref.~\cite{Taurus}. It consists of regions (subclouds) with different volume particle density ranging from 100 to 1000 cm$^{-3}$. The average gas density in such subclouds is about $n_{H_2} = 300\mbox{ cm}^{-3}$, and the typical size is $d_{\rm cl} = 2.5$ pc. The average column density is $N_{H_2} = 2.1\times 10^{21}\mbox{ cm}^{-2}$. Note the approximate relation between the volume and column densities, $N_{H_2}\approx d_{\rm cl} n_{H_2}$.

In this section, we will study thermal radiation from a region (subcloud) of the Taurus molecular cloud with the typical size $d_{\rm cl} = 2.5$ pc and volume  density $n_{H_2} = 300\mbox{ cm}^{-3}$, rather than considering the full Taurus cloud. Indeed, radiation spectrum from different regions of the Taurus cloud may vary significantly, as the density in such subclouds may differ by 10 times. For simplicity, we will assume that this subcloud has spherical shape with radius $r_{\rm cl}=d_{\rm cl}/2$ and volume $V_{\rm cl} = \frac43\pi r_{\rm cl}^3$. The visible solid angle of such subcloud is $\Omega = 3\times 10^{-4}$ sr. Note that all these parameters are very approximate. Therefore, the accuracy of our estimate of the radiation will be within order of magnitude. However, this will be sufficient for the goals of this paper, as we aim to estimate the feasibility of observing this effect.

The radiation power from the cloud is proportional to the dark matter particle number density in the cloud. Recall that the dark matter density in the Sun neighbourhood is usually estimated as $\rho_{\rm DM} = 0.3$ GeV/cm$^3$. Assuming that the mass of one QN is on the order of $m_{\rm QN} = B$~GeV, the QN particle number density is 
\begin{equation}
\label{nDM}
n_{\rm DM} = \rho_{\rm DM}/m_{\rm QN} = 0.3 B^{-1}\mbox{cm}^{-3}\,.    
\end{equation}
As a result, we estimate the total number of DM particles in the cloud as
\begin{equation}
\label{numberQN}
    {\cal N}_{\rm DM} = V_{\rm cl} n_{\rm DM} = 8.7\times 10^{55}B^{-1}\,.
\end{equation}
To find the total radiation power from (anti)QNs in the molecular cloud we multiply the radiation power of one quark nugget (\ref{P1}) by (\ref{numberQN}):
\begin{equation}
    {\cal N}_{\rm DM} P_1 = 4\pi^2 R^2 V_{\rm cl} n_{\rm DM}E(\omega) I_0(\omega,T)\,.
    \label{P13}
\end{equation}
However, this estimate does not take into account the extinction of the light on the dust particles in the cloud. 

The extinction of radiation in the Taurus molecular cloud was studied in Ref.~\cite{TaurusExtinction}. It is described by the function $A_\lambda$. Some values of this function are given in Table \ref{tab:extinction}. Note that the radiation extinction $A_\lambda$ is related to the optical depth $\tau$ as $A_\lambda = 1.086\tau$. The latter may be written as $\tau = \sigma_{\rm ext} N_{\rm dust}$, where $\sigma_{\rm ext}$ is the radiation extinction cross section on dist particles and $N_{\rm dust}$ is the dust column density. Given the approximate relation $N_{\rm dust}= d_{\rm cl}n_{\rm dust}$, we find the radiation attenuation lengths in the molecular cloud:
\begin{equation}
\label{labs}
    l_{\rm abs} = \frac1{n_{\rm dust}\sigma_{\rm ext}} = \frac{d_{\rm cl}}{\tau}=\frac{1.086 d_{\rm cl}}{A_\lambda}\,.
\end{equation}
The values of the attenuation length for some wavelengths of interest are given in Table~\ref{tab:extinction}.

\begin{table}[tb]
    \centering
    \begin{tabular}{c|c|c|c|c}\hline\hline
        $\omega$, eV & 0.5 & 1 &2 &3 \\\hline
        $\lambda^{-1},\ \mu{\rm m}^{-1}$  & 0.40 & 0.81 & 1.6 & 2.4 \\\hline
        $A_\lambda/A_V$ & 0.1 & 0.3 & 0.8 & 1.3 \\\hline
        $l_{\rm abs}$, pc & 4.8 & 1.6 & 0.6 &0.4 \\\hline
        $V_{\rm eff}/V_{\rm cl}$ & 1 & 0.83 & 0.35 & 0.24 \\\hline
        $P_{\rm obs},\ \frac{10^{-21}}{B^{1/3}}\frac{\mbox{ erg}}{\mbox{s Hz cm}^2}$ & 1.2 & 3.4 & 2.8 & 0.86 \\\hline\hline
    \end{tabular}
    \caption{Values of radiation extinction $A_\lambda$ for Taurus molecular cloud derived from Ref.~\cite{TaurusExtinction} with $A_V=5.7$. Here $l_{\rm abs}$ is the radiation attenuation length, $V_{\rm eff}$ is the effective cloud volume responsible for the radiation outside the cloud, and $P_{\rm obs}$ is observable energy flux on Earth.}
    \label{tab:extinction}
\end{table}

The radiation attenuation length (\ref{labs}) effectively reduces the volume of the molecular cloud which is responsible for possibly observable radiation outside the cloud. Assuming that the radiation from a single QN inside the cloud is limited by the sphere of radius $l_{\rm abs}$, the effective volume of the cloud is given by $V_{\rm eff} = \pi l_{\rm abs}(r_{\rm cl}^2 - l^2_{\rm abs}/12)$. As the attenuation length depends on radiation wavelength, the effective volume of the cloud reduces for high frequencies. Numerical values of $V_{\rm eff}$ for the wavelengths of interest are given in Table~\ref{tab:extinction}.

Replacing the cloud volume $V_{\rm cl}$ with the effective volume $V_{\rm eff}$ in Eq.~(\ref{P13}), we find the total thermal radiation power from anti-QNs in Taurus molecular cloud:
\begin{equation}
    P_{\rm tot}(\omega) = 4\pi^2 \mbox{fm}^2 \mbox{GeV}^{-1} B^{-1/3} V_{\rm eff} \rho_{\rm DM}E(\omega) I_0(\omega,T)\,.
    \label{Ptot-result}
\end{equation}
Note that this formula describes the radiation power in the position of the cloud in all directions. To obtain the radiation flux from this cloud measured on Earth, we divide Eq.~(\ref{Ptot-result}) by the area of the sphere with radius $L=140$ pc:
\begin{equation}
    P_{\rm obs}(\omega) = \frac{P_{\rm tot}(\omega)}{4\pi L^2}=\frac{\pi \,\mbox{fm}^2}{B^{1/3}L^2 {\rm GeV}}V_{\rm eff} \rho_{\rm DM} E(\omega) I_0(\omega,T) \,.
\label{Pobs}
\end{equation}
Numerical values of this function for some frequencies of interest are collected in Table~\ref{tab:extinction}.

For a comparison, let us consider the observable energy flux (\ref{Pobs}) for a typical value of the baryon charge number in the QN model, $B=10^{24}$ (see, e.g., \cite{Zhitnitsky-review}). For the frequency $\omega = 2.23$ eV ($\lambda = 555$ nm, visible V-band), the radiation power is
\begin{equation}
    P_{\rm obs} = 1.2\times 10^{-29}\frac{\rm erg}{\mbox{s Hz cm}^2}\,.
\end{equation}
This corresponds to the following visible and absolute AB magnitudes:
\begin{eqnarray}
    m_{\rm AB}&=& -2.5 \log_{10}( P_{\rm obs}) -48.6 = 23.2\,,\\
    M_{\rm AB} &=& m_{\rm AB}-5 \log_{10}L +5 = 17.5\,.
\end{eqnarray}

For a comparison, recall that the absolute AB magnitude of the Sun in V band is $M_{\rm AB, sun} = 4.83$. Thus, the molecular clouds are very faint sources of light. However, the light from anti-QN annihilation in molecular clouds could be registered by modern telescopes. Indeed, the faintest object observed by Hubble telescope has apparent magnitude $m=31.5$, which corresponds to the flux $P=9\times 10^{-33}\mbox{erg s}^{-1}\mbox{Hz}^{-1}\mbox{cm}^{-2}$. Thus, the radiation from cold molecular cloud could be registered for $B<2\times 10^{33}$. The non-observation of this radiation, instead, pushes the parameter $B$ beyond this limit.
 
Note that here we considered the radiation from regions of the Taurus molecular cloud with an average gas density $n_{\rm H_2} = 300$ cm$^{-3}$. In a similar way it is possible to estimate the radiation from denser regions with $n_{\rm H_2} \sim 1000$ cm$^{-3}$. Because of the higher gas density, the effective anti-QN temperature is slightly higher, $T=0.7$ eV. The corresponding radiation power in the visible V-band is $P_{\rm obs} = 6.9\times 10^{-29}\mbox{erg s}^{-1}\mbox{Hz}^{-1}\mbox{cm}^{-2}$. Such subclouds should be brighter, with visible and absolute AB magnitudes $m_{\rm AB} = 21.8$ and $M_{\rm AB} = 16.1$, respectively. Such subclouds could be visible with the Hubble space telescope for $B<4\times 10^{35}$.


\section{Non-thermal radiation from molecular cloud}
\label{sec-nonthermal}

There are three main types of non-thermal radiation from quark nuggets:
\begin{itemize}
    \item[A.] The $\gamma$ photons in the range $\sim$100-400 MeV produced by $\pi^0$ mesons originating from the hydrogen annihilation;
    \item[B.] The 511 keV line of positron annihilation evaporated from the positron cloud after hydrogen annihilation;
    \item[C.] The MHz-range synchrotron radiation from ultrarelativistic electrons and positrons produced by $\pi^\pm$ from the hydrogen annihilation.
\end{itemize}
In this section we estimate these three types of radiation for Taurus cloud.

\subsection{$\gamma$ photons}

On average, each hydrogen annihilation event on the anti-QN yields two $\pi^0$ mesons, each has the main decay channel into two $\gamma$ photons with an energy of order $\sim200$ MeV. Therefore, first, we have to estimate the hydrogen annihilation rate $W$ in Taurus molecular cloud. 

Let $\sigma=\pi R^2=\pi\, {\rm fm}^2B^{2/3}$ be the QN annihilation cross section and $v=10^{-3}c$ the velocity of QN particles in the galaxy. The hydrogen annihilation rate is
\begin{equation}
    W = \sigma v n_b n_{\rm DM}\,,
\end{equation}
where $n_b=600\mbox{ cm}^{-3}$ is the baryon density in the molecular cloud and $n_{\rm DM }$ is the DM particle number density in the cloud (\ref{nDM}). With these parameters, we find
\begin{equation}
\label{Wgamma}
    W = 2\times 10^{-16} B^{-1/3} \mbox{ s}^{-1} \mbox{cm}^{-3}\,.
\end{equation}

Each hydrogen annihilation event produces on average $N\simeq4$ gamma photons. Roughly, these photons have mean energy ${\cal E}\sim200$ MeV in the interval between 100 and 300 MeV, that corresponds to the spectral density $dN/d{\cal E} \sim 0.02\mbox{ MeV}^{-1}$. Multiplying this spectral density by the hydrogen annihilation rate (\ref{Wgamma}), we estimate the photon production rate in the molecular cloud:  
\begin{equation}
    F = V_{\rm cl}  \frac{dN}{d{\cal E}} W = 7.7\times 10^{39} B^{-1/3} \mbox{ MeV}^{-1} \mbox{ s}^{-1}\,.
\end{equation}
The flux of observed photons on the Earth is
\begin{equation}
    \Phi_{\rm QN} = \frac{F}{4\pi L^2} = 3.3\times 10^{-3} B^{-1/3} \mbox{ MeV}^{-1} \mbox{ s}^{-1}\mbox{ cm}^{-2}\,.
\end{equation}
Given that the visible solid angle of this subcloud in the Taurus molecular cloud is $\Omega = 3\times 10^{-4}$ sr, the observed flux per unit solid angle is
\begin{equation}
    \Phi_{\rm QN}\Omega^{-1} = 11 B^{-1/3}\mbox{ MeV}^{-1} \mbox{ s}^{-1}\mbox{ cm}^{-2}\mbox{sr}^{-1}\,.
    \label{21}
\end{equation}
For $B=10^{24}$, we have
\begin{equation}
    \Phi_{\rm QN}\Omega^{-1} = 1.1 \times 10^{-7}\mbox{ MeV}^{-1} \mbox{ s}^{-1}\mbox{ cm}^{-2}\mbox{sr}^{-1}\,.
    \label{22}
\end{equation}

The estimated $\gamma$ photon flux from anti-QNs should be compared with the one measured by the Fermi-LAT telescope \cite{Clouds-gamma}: at ${\cal E}=300$ Mev, the flux is $\Phi(300\mbox{ MeV})\approx 1.5 \Phi(3\mbox{ GeV})$, where $\Phi(3\mbox{ GeV})=9.8\times 10^{-9}\mbox{ GeV}^{-1}\mbox{s}^{-1}\mbox{cm}^{-2}$. This flux is measured from the region on the sky with the angular area $\Omega = 14^\circ\times 14^\circ=0.06$ sr. Thus, the measured flux per steradian is 
\begin{equation}
    \Phi(300\mbox{ MeV})\Omega^{-1} = 2.5\times 10^{-10} \mbox{ MeV}^{-1}\mbox{s}^{-1}\mbox{cm}^{-2} \mbox{sr}^{-1}\,.
    \label{gamma-observed}
\end{equation}

Our estimate of photon flux from anti-QNs for  $B=10^{24}$ (\ref{22}) exceeds the photon flux detected by the Fermi telescope \cite{Clouds-gamma}. Formally, the predicted flux (\ref{22}) is consistent with the observed one (\ref{gamma-observed}) for $B\geq 8.5 \times 10^{31}$. However, this does not allow us to exclude the values of the baryon charge below this value, because different parts of the Taurus cloud are considered here. Indeed, in our estimate we considered a small subcloud in the Taurus cloud of size about 2.5 pc, while in Ref.~\cite{Clouds-gamma} a large region of the size $\sim11$ pc was considered. The average gas density in the considered ragions may be significantly different. Thus, from this comparison we can only conclude that {\it it is plausible that annihilation of anti-QNs in giant molecular clouds produces an observable flux of $\gamma$ photons}, while for a quantitative conclusion a more accurate treatment is needed. 

\subsection{511 keV line} 

Each hydrogen annihilation event reduces the QN core electric charge and rises the effective temperature of the positron cloud. Thus, at least one positron evaporates from this positron cloud as a result of the hydrogen annihilation. This positron should eventually annihilate with the hydrogen gas in the interstellar medium. This annihilation goes through the formation of a positronium state with subsequent decay. One quarter of these states correspond to the para-positronium which decay into two 511 keV photons. Thus, the 511 keV photon production rate in the molecular cloud may be roughly estimated as
\begin{equation}
    F = \frac12 V_{\rm cl} W = 0.2\times 10^{42} B^{-1/3} \mbox{ s}^{-1}\,,
\end{equation}
where $W$ is given by (\ref{Wgamma}).

The flux of observed photons on the Earth is
\begin{equation}
    \Phi_{\rm QN} = \frac{F}{4\pi L^2} = 0.085 B^{-1/3} \mbox{ s}^{-1}\mbox{ cm}^{-2}\,.
\end{equation}
This must be compared with the sensitivity of the SPI/INTEGRAL detector to the 511 keV line: 
\begin{equation}
    \label{sensitivity}
    \Phi_{\rm sensitivity} = 5\times 10^{-5} \mbox{ s}^{-1} \mbox{cm}^{-2}\,.
\end{equation}
Thus, the radiation from QNs in the molecular cloud could be observed by the SPI/INTEGRAL detector if
\begin{equation}
    B<5\times 10^9\,.
\end{equation}
Taking into account that there are areas in the cloud with a higher matter density and density of dark matter locally may be higher,  this number   for $B$ may be bigger. However, we should conclude that for $B>10^{24}$ the  511 keV photons from anti-QN annihilation in the cold hydrogen clouds  can hardly be detected. 

\subsection{Synchrotron radiation}

Each hydrogen annihilation event produces on average two charged $\pi$ mesons near the QN boundary. Mean energy of these pions is on the order of 400 MeV. One of these pions goes inside the QN core, termalize and further decay. The other meson goes outwards and escapes. The final decay product of this meson is either an electron or positron and neutrinos. These electrons (and positrons) are ultrarelativistic with energy of order 400 MeV. In the cloud, they move along the lines of weak magnetic field $H\sim 10\mu$G and are the source of synchrotron radiation.

The characteristic time of ultrarelativistic electrons and positrons in the cloud may be estimated as $t\sim R_{\rm cl}/c\approx 8$ y. Thus, the number of such electrons and positrons in any moment of time is estimated as
\begin{equation}
\label{electrons-in-cloud}
    N = V_{\rm cl} W t =10^{50} B^{-1/3}\,,
\end{equation}
where $W$ is given by (\ref{Wgamma}).

The spectral density of synchrotron radiation is given by \cite{LL2}
\begin{equation}
    I(\omega) = \frac{\sqrt3}{2\pi} \frac{e^3 H}{mc^2} F(\omega/\omega_c)\,,
\end{equation}
where 
\begin{equation}
\label{Fx}
    F(x) = x\int_x^\infty K_{5/3}(y)dy\,,\qquad
    \omega_c = \frac{3eH}{2mc}\left(\frac{\cal E}{mc^2} \right)^2\,.
\end{equation}
Here $K_{5/3}$ is the modified Bessel function of the second kind and ${\cal E}=400$ MeV is the particle's energy. In the magnetic field $H=10\,\mu$G, the electron has the frequency $\omega_c\approx 10^{-7}\mbox{ eV}=170$ MHz. The function (\ref{Fx}) reaches its maximum $F_{\rm max} = 0.92$ at $\omega = 0.29 \omega_c = 44$ MHz. For this frequency, we find
\begin{equation}
    I_{\rm max} = \frac{\sqrt3 e^3 H}{2\pi mc^2}F_{\rm max} \approx 3.4\times 10^{-28} \mbox{ erg s}^{-1}\mbox{Hz}^{-1}\,.
\end{equation}
The spectral density from all ultrarelativistic electrons and positrons in the cloud is 
\begin{equation}
    N I_{\rm max} = 3.4\times 10^{22} B^{-1/3} \mbox{ erg s}^{-1}\mbox{Hz}^{-1}\,.
\end{equation}

The corresponding radiation power on the Earth is
\begin{equation}
    P_{\rm QN} = \frac{N I_{\rm max}}{\Omega S} = 5.8\times 10^{-17} B^{-1/3} \frac{\rm erg}{\mbox{Hz sr s cm}^2}\,,
\end{equation}
where $\Omega= 3\times 10^{-4}$ sr is the visible solid angle of the considered region in Taurus molecular cloud and $S = 4\pi L^2$ is the area of the sphere, $L=140$~pc is the distance to the cloud. Unfortunately, this radiation power cannot be resolved from the background \cite{RF1,RF} $P_{\rm bg}\sim 3\times 10^{-18}\mbox{erg s}^{-1}\mbox{cm}^{-2}\mbox{Hz}^{-1}\mbox{sr}^{-1}$.





\section{Can anti-quark nuggets  explain the solar corona temperature mystery?}
\label{sec-corona}

The papers \cite{corona1,corona2} address the issue of the solar corona temperature within the AQN framework. The authors of these papers claim that the annihilation of anti-quark nuggets in the solar corona plasma can release enough energy to raise its temperature to about million K. In this section, we will show that this conclusion is not justified.

\subsection{Total energy from all quark nuggets in the solar corona}
\label{sec-corona-total}

It is known (see, e.g.,\cite{CoronaEmission}) that the total radiation from the solar corona is on the order of 
\begin{equation}
\label{Fcorona}
F_{\rm corona} = (1-5)\times 10^{27}\mbox{erg s}^{-1}\,.
\end{equation}
In Refs.~\cite{corona1,corona2}, it was argued that similar energy flux could be produced by dark matter particles completely ($\sim$ 100\%) annihilating inside the solar corona.

Denote by $\xi=\delta m_{\rm QN}/m_{\rm QN}$ the relative mass loss of an anti-QN due to the annihilation in the solar corona. Then, the total energy flux produced by all anti-QNs annihilating in the solar corona is
\begin{equation}
\label{FQN}
    F_{\rm QN} = 4\pi R_{\odot}^2\xi \gamma_\odot v \rho_{\rm DM}=4.6\times 10^{27} \xi \mbox{ erg s}^{-1}\,,
\end{equation}
where $v=10^{-3}$ is the typical velocity of DM particles, $\rho_{\rm DM} = 0.3 \mbox{ GeV cm}^{-3}$ is the dark matter density in the Sun neighborhood, and $\gamma_\odot$ is the enhancement factor due to the gravitational attraction. By definition, this factor relates the effective capture cross section of the Sun with its geometric cross section, $\sigma_{\rm eff} = \gamma_\odot \pi R_\odot^2$. It is determined by the classical energy and angular momentum conservation conditions of a falling body,
\begin{equation}
 \gamma_\odot = 1+ \frac{2GM_\odot}{R_\odot v^2} \approx 5.2\,.
\label{gamma-sun}
\end{equation} 
Comparing Eqs.~(\ref{Fcorona}) and (\ref{FQN}) the authors of Refs.~\cite{corona1,corona2} concluded that the anti-QNs should annihilate completely in the solar corona, $\xi\simeq 1$, in order to fully explain the radiation from the solar corona. In this section, we will show that the density of the solar corona is so low that it cannot annihilate a significant portion of anti-QN's mass even with the most optimistic assumptions about the annihilation cross section.

The parameters of density, temperature and width of the solar corona slightly vary in different studies. Here we will use the values of these parameters adopted in Ref.~\cite{corona2}. In particular, we assume that the deepest region of the solar corona has gas density $n=10^{10}\mbox{ cm}^{-3}$ and temperature $10^6$ K. Although the solar corona extends thousands kilometers above the Sun surface, it is sufficient to consider the lowest layer of the depth about 3000 km, as it gives the main contribution to the annihilation of anti-QNs. To make an upper estimate of the radiation from annihilation events in the solar corona we assume the highest value $n=10^{10}\mbox{ cm}^{-3}$ in the layer of 3000 km above the chromosphere. A similar assumption was used in the estimates in Ref.~\cite{corona2}. 

\subsection{Annihilation of one anti-QN in the solar corona}

The QN particle approaches the Sun from large distance with an average velocity $v=10^{-3}c$. 
Near the Sun surface its velocity would be roughly $v=2\times 10^{-3}c$ due to the gravitational attraction. Therefore, following Ref.~\cite{corona2}, we set the initial velocity $v=2\times 10^{-3}c$ for anti-QNs at the altitude $h=3000$ km. 

Let us consider first the most simple case of an anti-QN in a head-on collision with the Sun. Classically, its motion is determined by the gravitational attraction and collisional friction forces. It may be shown that the former is at least twelve orders in magnitude stronger than the latter, see Eqs.~(\ref{Fgrav}) and (\ref{Fstop}) in Appendix. Hence, the collisional friction may be neglected. It is easy to estimate the acceleration due to the gravitational attraction of the anti-QN when it crosses the lowest 3000 km thick layer of the solar corona: $\delta v/v = 2\times 10^{-3}$. Thus, to a good accuracy, we can consider the anti-QN moving through the solar corona with the constant velocity $v = 2\times 10^{-3}c$. With this velocity, the anti-QN particle needs about 5 seconds to cross this layer in the solar corona, before it sinks in the denser regions of the chromosphere and photosphere.

The number of collisions of the anti-QN with the gas particles in the solar corona is estimated as
\begin{equation}
    N_{\rm col} =n \sigma_{\rm col} h \,,
    \label{Ncol-corona}
\end{equation}
where $\sigma_{\rm col}$ is the anti-QN collision cross section with protons in the plasma. It differs from the geometric cross section $\sigma= \pi R^2$ by an enhancement factor $\gamma$, $\sigma_{\rm col} = \gamma \pi R^2$. Recall that this factor takes into account the Coulomb attraction potential $U(r)$ of protons to the charged anti-QN, because the latter is partly ionized in the hot plasma. Since anti-QN is a macroscopic particle with characteristic size $10^{-5}$ cm, the enhancement factor may be found from classical conditions of energy and angular momentum conservation of a falling body, by analogy with Eq.~(\ref{gamma-sun}), 
\begin{equation}
    \gamma = 1 + \frac{|U(R)|}{{\cal E}_{{\rm kin},\infty}}\,,
    \label{gamma-factor}
\end{equation}
where $|U(R)| = \frac{Qe}{R}$ is the Coulomb potential of anti-QN with charge $Q$ near its surface, and ${\cal E}_{{\rm kin},\infty} = \frac{m_p v^2}{2}$ is the kinetic energy of a proton far away from the anti-QN.

According the the virial theorem, the ionization potential $I$ is equal to one half of the average potential energy  of bound  positrons, orbiting anti-quark nugget,  $I=|U(R)/2|$. If the  temperature of the positron gas  $T$ is higher than the ionisation potential, the positron will leave the  nugget. Therefore, we assume $I=|U(R)/2|=k_BT$. 

The temperature $T$ may be estimated in two different ways. First, the temperature rises due to the incoming energy from proton annihilation on anti-QNs. Using the methods developed in Ref.~\cite{FS2}, we find this temperature $T\simeq 24$ eV. Second, if the anti-QN is in thermal equilibrium with the surrounding plasma, this temperature should be of order $T=100$ eV that corresponds to the corona temperature $10^6$ K $\sim 100$ eV. We will assume the latter temperature for the anti-QN to find the upper estimate in this section, although it is unlikely that anti-QNs set up thermal equilibrium with the radiation in the solar corona so quickly. Thus, for the mean potential energy in the positron cloud we take $|U(R)|=200$ eV. Substituting this value into Eq.~(\ref{gamma-factor}) together with the kinetic energy of protons impacting the anti-QN with the velocity $v= 2\times 10^{-3}c$, we find $\gamma = 1.1$. Note that this is an upper estimate for the enhancement factor, as we have not taken into account the screening effects. Note also that the value of this enhancement factor was strongly overestimated in Ref.~\cite{corona2}, $\gamma\sim 10^8$. This large value of the enhancement factor corresponds to the elastic scattering of protons off anti-QN rather than their capture and annihilation. In Appendix \ref{AppA}, we demonstrate that the proton scattering plays minor role in anti-QN dynamics in the solar corona and, thus, it may be neglected.
 
In the solar corona, the anti-QNs are only partly ionized. The interaction of the impacting proton with the positron cloud does not allow the former to elastically re-bounce from the anti-QN surface, as there is energy loss due to the friction resulting in proton capture \cite{FS21}. Therefore, the chance of annihilation of the incident proton on the anti-QN is close to one. Therefore, making use of Eq.~(\ref{Ncol-corona}), we write the number of proton annihilations for the anti-QN in the solar corona:
\begin{equation}
N_{\rm ann} \approx N_{\rm col} =  \pi R^2 n \gamma h = 10^{-7} B^{2/3}\,,
\end{equation}
where we made use of the relation $R=B^{1/3}{\rm fm}$. Since each annihilation reduces the baryon number by one, the anti-QN mass loss is
\begin{equation}
    \xi = \frac{\delta m_{\rm QN}}{m_{\rm QN}} = \frac{N_{\rm ann}}{B} = 10^{-7} B^{-1/3}\,.
    \label{xi}
\end{equation} 
For $B > 10^{24}$ we obtain  $\xi <10^{-15}$, and the energy flux from anti-QNs (\ref{FQN}) is 15 orders of magnitude lower than the total radiation from the solar corona (\ref{Fcorona}). Thus, we conclude that {\it the anti-QNs cannot be responsible for the solar corona heating}. 

In the above estimate, following Ref.~\cite{corona2}, we assumed that the quark nugget passes the length $h=3000$ km in the solar corona.  The path of tangentially moving anti-QNs is about 20 times longer than that for the anti-QNs in the head-on collision. However, this does not change the conclusion $\xi\ll1$, as the expression (\ref{xi}) indicates a very strong suppression factor.


\section{Radiation from chromosphere}
\label{sec-chromosphere}

As is demonstrated above, the anti-QN annihilations in the solar corona are relatively rare since the gas density is low there. In this section, we consider the radiation from anti-QNs annihilating in the chromosphere, where the gas density is much higher. This radiation could be observable as the chromosphere is transparent for most of the radiation.

\subsection{Annihilation of anti-QNs in chromosphere}

The width of the chromosphere is about $h=2000$ km. The gas density in chromosphere is approximately described by the function 
\begin{equation}
n(z) = n_0 e^{-z/z_0}\,,
\label{n-chrom}
\end{equation}
where $z$ is the altitude above the photosphere, $n_0 = 10^{17}\mbox{ cm}^{-3}$ and $z_0 = 350$ km. These parameters are derived from Ref.~\cite{ChromoStructure} with graphical accuracy.

Let us first consider a trajectory of an anti-QN moving in the chromosphere normal to its surface. Along this trajectory the anti-QN experiences the following number of collisions:
\begin{equation}
    \pi R^2 \int_0^h n(z)dz =0.05 B^{2/3}\,,
\label{Nnormal}
\end{equation}
where we made use of the identity $R= B^{1/3}{\rm fm}$. This calculation, however, strongly underestimates the number of collisions for an average anti-QN impacting the Sun, as one has to consider a family of trajectories with various angles to the surface. It is possible to show that averaging over all these trajectories brings a factor of about 10 to the estimate (\ref{Nnormal}):
\begin{equation}
    N_{\rm col} \simeq 0.5 B^{2/3}\,.
\label{Ncol-cromo}
\end{equation}


Recalling that the number of annihilation is $N_{\rm ann} = \kappa N_{\rm col}$, we find the anti-QN mass loss during its motion in the chromosphere,
\begin{equation}
 \xi = \frac{\delta m_{\rm QN}}{m_{\rm QN}} = 0.5 \kappa B^{-1/3}\,.
\end{equation}
This number is significantly larger than  that in the solar corona (\ref{xi}).

\subsection{Temperature and spectrum of anti-QNs in chromosphere}

When the anti-QN moves through the medium, the annihilation events rise the effective temperature in the positron cloud and partly ionize it. Each nucleon annihilation brings about 1 GeV of energy in the positron cloud. In equilibrium, this energy is radiated through the thermal radiation with the power (\ref{P1}). 

Denote $W_{\rm out}(T) = \int_0^\infty P_1(\omega,T)d\omega$ the total radiation power of one anti-QN. It measures the outgoing energy flux through the surface of the anti-QN. In equilibrium, this flux is equal to the flux of incoming energy due to the hydrogen annihilation,
\begin{equation}
    W_{\rm in}(n) = 1\mbox{ GeV}\, \sigma_{\rm ann} v n\,,
\end{equation}
where $\sigma_{\rm ann}\approx \pi R^2$ is the anti-QN annihilation cross section, $v$ is the anti-QN velocity in the medium and $n$ is the nucleon number density of the medium. Thus, the effective temperature of anti-QNs is determined by the equation
\begin{equation}
    W_{\rm out}(T) = W_{\rm in}(n)\,.
    \label{equilibrium}
\end{equation}

\begin{figure}[tb]
    \centering
    \includegraphics[width=8cm]{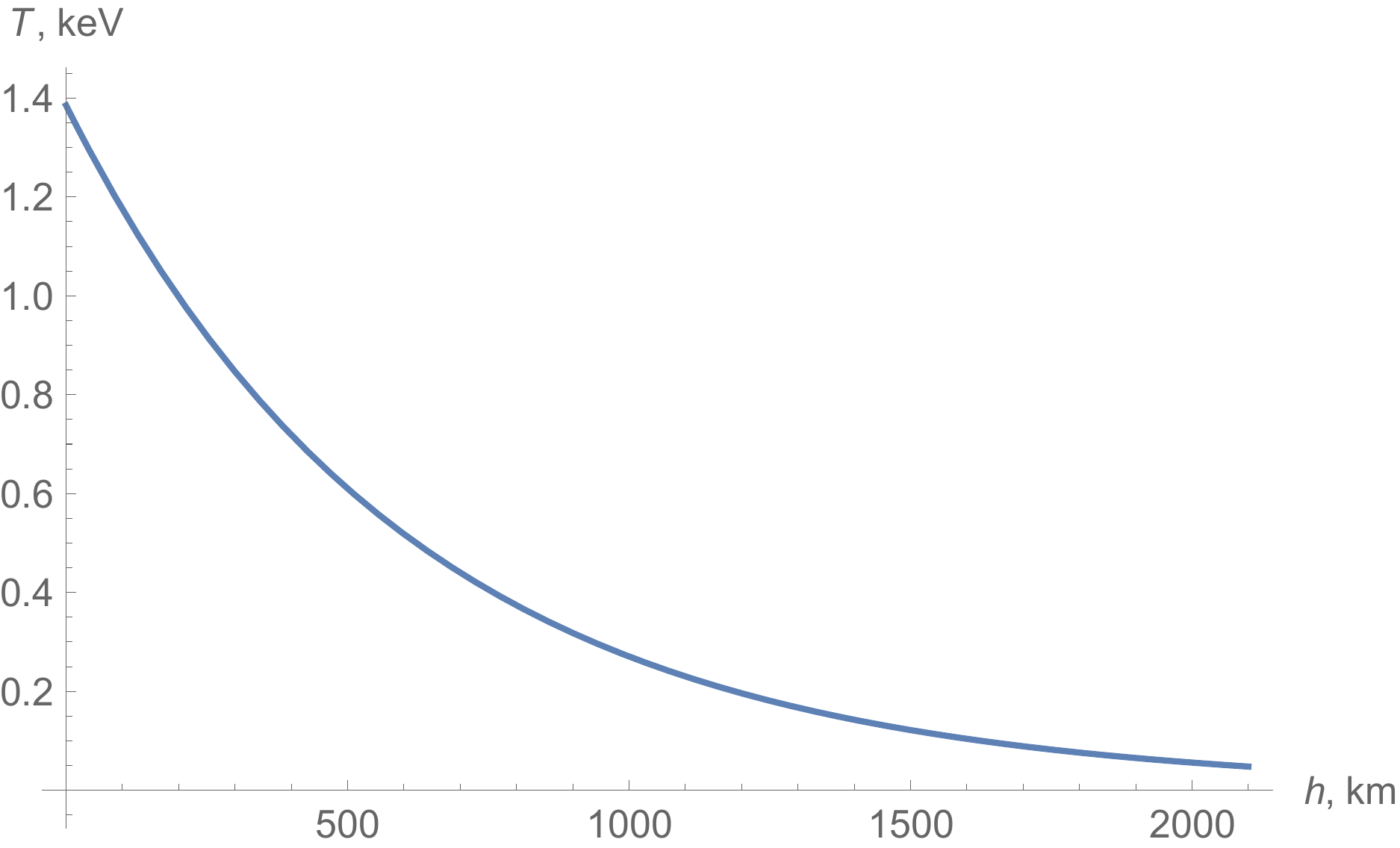}
    \caption{Effective anti-QN temperature in chromosphere as a function of the altitude $h$ above the photosphere.}
    \label{fig-temp}
\end{figure}

Equation (\ref{equilibrium}) may be solved numerically for each particular value of the density of chromosphere $n$. In particular, in the bottom of the chromosphere, where the density is the highest, $n_{\rm high} \simeq 10^{17}\mbox{ cm}^{-3}$, the effective anti-QN temperature is very high, $T_{\rm high} = 1.4$ keV. At the top of the chromosphere, where the gas density is of order $n=10^{11}\mbox{ cm}^{-3}$, the effective temperature is $T_{\rm low} = 50$ eV. The dependence of the effective temperature on the altitude $h$ above the photosphere is shown in Fig.~\ref{fig-temp}.

\begin{figure*}[tbh]
    \centering
    \includegraphics[width=16cm]{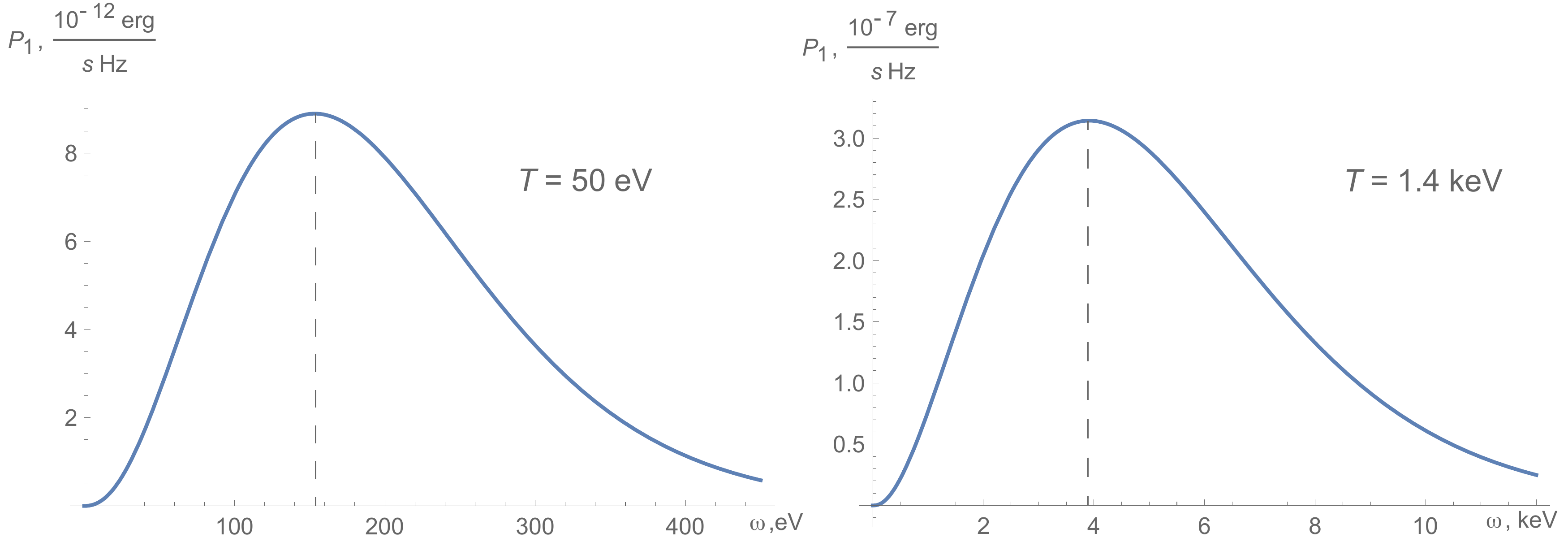}
    \caption{Spectrum of thermal radiation of anti-QNs at different temperatures.}
    \label{fig-spectrum}
\end{figure*}

The spectrum of thermal radiation of anti-QNs (\ref{P1}) strongly depends on its effective temperature. In Fig.~\ref{fig-spectrum}, we present the plots of the spectral density at the lowest and highest anti-QN temperature in the chromosphere. These figures show that the maximum of thermal radiation from anti-QNs in chromosphere varies from $\omega = 150$ eV to 4 keV. Thus, the radiation from anti-QNs in the chromosphere ranges from hard UV to X-rays.

However, the absorption of the UV and x-ray radiation in the chromosphere is very effective. Indeed, the photo ionisation cross section for 1$s$ electron in the Born approximation $\mbox{13.6 eV}\ll \hbar \omega \ll m_e c^2$ is (see, e.g., \cite{LL4})
\begin{equation}
\label{sigma-ion}
\sigma(\omega)=  \frac{2^8 \pi}{3} \alpha a_B^2 Z^5 \left(\frac {      {\cal E}_0    }{\hbar\omega}\right)^{7/2}\,,
\end{equation}
where ${\cal E}_0=e^4 m_e /2\hbar^2 = 13.6$ eV is the hydrogen atom ionization energy. Here $Z=1$ for hydrogen and $Z=2-5/16=1.69$ for helium (see, e.g., \cite{LL3}). Note also that for helium the cross section (\ref{sigma-ion}) should be multiplied by 2 to take into account for both electrons. Taken also into account that helium abundance in the chromosphere is about 8\%, we find the photon absorption in the chromosphere
\begin{equation}
    \sigma_{\rm ch}(\omega) \approx  2^8 \pi \alpha a_B^2 \left(\frac {      {\cal E}_0    }{\hbar\omega}\right)^{7/2}\,.
\end{equation}

\begin{figure}[tb]
    \centering
    \includegraphics[width=8cm]{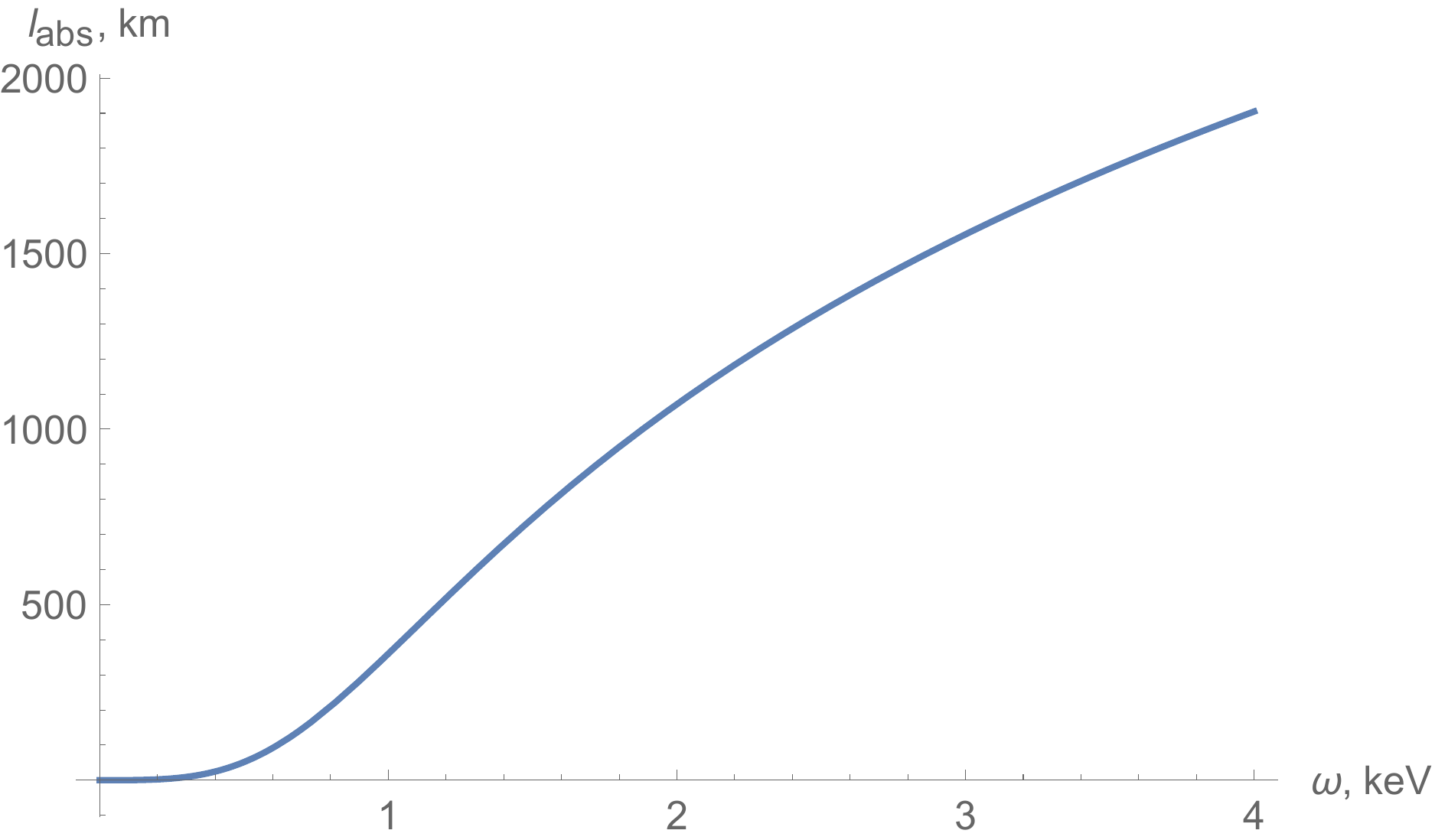}
    \caption{Photon attenuation length in chromosphere as a function of photon energy.}
    \label{fig-absotprion}
\end{figure}

The photon attenuation length in chromosphere may be found from the Beer–Lambert law, $I(x)=I_0 e^{-n(x)\sigma_{\rm ch}x}$, where $x=h-z$ is the distance from the top of the chromosphere, and the gas density $n$ is modelled by the function (\ref{n-chrom}). Considering that this density reduces with the altitude, we find that the radiation intensity $I$ reduces $e$ times on the distance
\begin{equation}
    l_{\rm att}(\omega) = z_0 \ln\left( 1 + \frac{e^{h/z_0}}{n_0 z_0 \sigma_{\rm ch}(\omega)} \right)\,.
\end{equation}
Recall that $n_0=10^{17}\mbox{ cm}^{-3}$ and $z_0 = 350$ km. This photon attenuation length is plotted in Fig.~\ref{fig-absotprion}. One can see from this plot that the chromosphere absorbs most of the photons with with energies under $\omega = 4$ keV. As is seen from Fig.~\ref{fig-spectrum}, the thermal radiation from quark nuggets in chromosphere falls into this region. Thus, {\it chromosphere screens the thermal radiation from quark nuggets}, while it is transparent for frequencies above 4 keV.

\subsection{511 keV line and $\gamma$ photons}

Anti-QNs are sources of 511 keV and gamma photons which appear in annihilations of atoms of visible matter on anti-QNs \cite{FS2,FS21}. This radiation should be produced by anti-QNs moving through the chromosphere and photosphere of the Sun. As is shown in the previous subsection (see also Fig.~\ref{fig-absotprion}), the chromosphere is transparent for these frequencies. In this subsection we estimate the flux of these photons on the Earth. 

Let us first estimate the flux of 511 keV photons on Earth due to anti-QNs annihilations in the chromosphere. Each proton annihilation on the anti-QN reduces the electric charge of its core and makes at least one of the positrons from the positron cloud to evaporate. This positron will annihilate in the gas of the chromosphere with the x-ray emission. Note that the positron annihilation produces either three photons through the ortho-positronium state or two photons through the para-positronium. The latter state decays into two 511 keV photons, with branching ratio 1/4. Thus, each nucleon annihilation event on the anti-QN produces approximately 1/2 of 511 keV photons.

The photon produced in the chromosphere may be directed either inside or outside the Sun. The inward going photons will be absorbed while the out-going ones can leave the chromosphere with subsequent detection. Therefore, we have to divide by half the number of 511 keV photons from the Sun. 

The number of collisions of one anti-QN passing through the chromosphere is given by Eq.~(\ref{Ncol-cromo}). The corresponding number of annihilations is $N_{\rm ann}=\kappa N_{\rm col}$, with $\kappa\approx 1$. The flux of DM particles on the Sun is
\begin{equation}
F_{\rm DM} = 4\pi R_\odot^2 \gamma n_{\rm DM}v = 2.8\times 10^{30}B^{-1} \mbox{ s}^{-1}\,,
\label{FDM}
\end{equation}
where $v=10^{-3}c$ and $n_{\rm DM} = 0.3 B^{-1} \mbox{cm}^{-3}$ are the typical velocity of DM particles and their number density (\ref{nDM}) in the solar neighbourhood, respectively, and $\gamma=5.2$ is the enhancement factor (\ref{gamma-sun}). As a result, the total flux of 511 keV photons from the surface of the Sun is the product of Eqs.~(\ref{Ncol-cromo}) and (\ref{FDM}):
\begin{equation}
    F_{511} = \frac14 N_{\rm col} F_{\rm DM} = 3.6 \times 10^{29} B^{-1/3} \mbox{ s}^{-1}\,.
\end{equation}
Here the factor $\frac14$ takes into account the number of 511 keV photons from each nucleon annihilation and photons directed outwards the Sun surface as per discussion above. The  flux of these photons on the Earth is 
\begin{equation}
    \Phi_{511} = \frac{F_{511}}{4\pi \mbox{AU}^2} = 125 B^{-1/3} \mbox{ cm}^{-2}\mbox{s}^{-1}\,.
    \label{Phi511}
\end{equation}
This value may be compared with the observed flux of 511 keV photons from the Sun flares \cite{511}:
\begin{equation}
    \Phi_{511,\rm obs} = 0.08 \mbox{ cm}^{-2}{\rm s}^{-1}\,.
\end{equation}
Thus, annihilation of anti-QNs in the chromosphere may produce a comparable flux of 511 keV photons on the Earth for $B<3.5\times 10^9$. In a similar way, comparing Eq.~(\ref{Phi511}) with the sensitivity of the SPI/INTEGRAL detector (\ref{sensitivity}), we conclude that the predicted flux of 511 keV photons from the chromosphere could be observed if $B<1.6\times 10^{19}$. Thus, for the expected value of the baryon charge number $B>10^{24}$ the flux of 511 keV photons seems to be too small to be observed.

The consideration of $\gamma$ photons from anti-QN annihilation in the chromosphere goes along similar lines. We should only keep in mind that each hydrogen annihilation event on anti-QN yields on average two $\pi^0$ mesons which decay mainly into two $\gamma$ photons each. The energies of these photons mostly fall into the interval 100-400 MeV. Thus, each annihilation yields on average four $\gamma$ photons, two of which are radiated outside the Sun surface and may be detected on the Earth. The flux of such photons should be four times that of 511 keV photons (\ref{Phi511}):
\begin{equation}
    \Phi_\gamma = 4\Phi_{511} = 500 B^{-1/3} \mbox{ cm}^{-2} \mbox{s}^{-1}\,.
\end{equation}
Comparing this value with the one measured by Fermi-LAT \cite{Abdo_2011}
\begin{equation}
    \Phi_{\gamma,\rm obs} \simeq 4.6\times 10^{-7} \mbox{ cm}^{-2} \mbox{s}^{-1}\,,
\end{equation}
we conclude that this flux may be fully explained by the anti-QNs annihilations in chromosphere for $B<1.3\times 10^{27}$. Thus, for the expected value of $B>10^{24}$  the flux of 100-400 MeV photons from the anti-QN annihilation is sufficiently large  to be observed.


\section{Constraint from condition of survival of anti-QNs in early universe}
\label{sec-survival}

According to the scenario proposed in Ref.~\cite{AQNformation}, the quark nuggets in the early universe begin their formation in the QCD phase transition epoch, when the temperature of the universe was of order 160-170 MeV. In this process, the pion-axion domain walls effectively capture quarks and antiquarks and separate matter from antimatter. All dark matter is supposed to be represented by such quark and antiquark nuggets, while the visible matter corresponds to free baryons which escaped in this capture process. In this scenario \cite{AQNformation}, the QN and anti-QN formation process terminates at the temperature $T_{\rm form}=41$ MeV, when all domain walls either decay or form (anti)QNs. 

In contrast with QNs, the anti-QNs consist of antimatter, and, thus, may annihilate nucleons. Each collision of a nucleon  with the anti-QN reduces the baryon number of the latter with some probability. Thus, at the moment of formation, anti-QNs  should have a large enough baryon number to survive till the present day. As is shown in Ref.~\cite{Survival}, the anti-QNs can survive in the harsh environment of the early universe and be observed at the present day if they carry the baryon charge $B\gtrsim 10^{24}$. In this section, we will revisit this estimate.


The collision rate of QNs with baryons in the surrounding plasma is
\begin{equation}
\label{Gcol}
    \Gamma_{\rm col} = 4 \sigma_{\rm col} n_B v_B\,,
\end{equation}
where $\sigma_{\rm col}$ is the collision cross section, $n_B$ is the baryon density and $v_B$ is mean baryon velocity in any fixed direction. Let us consider these quantities carefully.

The collision QN cross section may be related to its geometric cross section $\sigma= \pi R^2$ as $\sigma_{\rm col} = \gamma \sigma$, where $\gamma$ is the Coulomb enhancement factor. For neutrons, this factor is obviously $\gamma_{n}=1$, while for protons $\gamma>1$. This factor may be estimated with the use of Eq.~(\ref{gamma-factor}) in which the potential energy is determined by the
QN ionization potential, $I = |U(R)/2|=k_B T$, and the kinetic energy is simply ${\cal E}_{{\rm kin},\infty} = \frac32 k_B T$. Substituting these values into Eq.~(\ref{gamma-factor}), we find
$ \gamma_p = 1+ \frac43 = \frac73$. Assuming equal number of protons and neutrons in the plasma, we find $ \gamma = \frac12(\gamma_p + \gamma_n) = \frac{5}{3}$. Thus, for the collision cross section in Eq.~(\ref{Gcol}) we obtain the following estimate
\begin{equation}
    \sigma_{\rm col} \approx  \frac53 \pi R^2\,.
\end{equation}

The baryon density $n_B$ is usually determined from the measured baryon-to-photon ratio (see, e.g., \cite{PDGreview})
\begin{equation}
\label{b/ph}
    \eta = \frac{n_B}{n_\gamma} = 6.15\times 10^{-10}\,,
\end{equation}
where 
\begin{equation}
    n_\gamma = \frac{2\zeta(3)}{\pi^2}T^3=0.244(k_B T)^3
\end{equation}
is the photon number density in CMB. However, the ratio (\ref{b/ph}) is measured in the CMB era while we need to use it in the pre electron-positron annihilation epoch. For this purpose, instead of (\ref{b/ph}) it is appropriate to use the ratio $n_B/s$ where $s$ is the total entropy density. In particular, in the current epoch, this entropy density is calculated as (see, e.g., \cite{PDGreview})
\begin{equation}
    s_{\rm now} = \frac43 \frac{\pi^2}{30} \frac{43}{11} (k_B T)^3 = 7.04 n_\gamma\,.
\end{equation}
In QN post formation epoch, electrons, neutrino and photons contributed as radiation. Therefore, the entropy density was
\begin{equation}
    s_{\rm form} = \frac43 \frac{\pi^2}{30} \frac{43}{4} (k_B T)^3\,.
\end{equation}
Thus, in the QN post formation epoch, the baryon-to-photon ratio (\ref{b/ph}) is enhanced by the factor $s_{\rm form}/s_{\rm now} = 11/4$,
\begin{equation}
\label{nB}
    n_{B} = \frac{11}4 \eta n_\gamma = 4.12\times 10^{-10} (k_BT)^3\,.
\end{equation}

Finally, Eq.~(\ref{Gcol}) involves the average velocity $v_B$ of baryons in plasma in a fixed direction. This velocity may be readily calculated by averaging $v_x$ with one-dimensional Boltzmann distribution function over the half-line,
\begin{equation}
\label{vB}
v_B\equiv \langle v_x \rangle = \left(\frac{k_B T}{2\pi{m_p}}\right)^{1/2}    \,.
\end{equation}

Substituting (\ref{vB}) and (\ref{nB}) into Eq.~(\ref{Gcol}), we find collision rate of an anti QN with nucleon in the primordial plasma
\begin{equation}
\label{GammaCol}
    \Gamma_{\rm col} = 3.4 \times 10^{-9} B^{2/3} {\rm fm}^2 T^{7/2} m_p^{-1/2}\,.
\end{equation}
Integrating this collision rate from the formation time $t_{\rm form}$ to the recombination time $t_{\rm rec}$, we get the total number of collisions of anti QNs with nucleons in the hot primordial plasma,
\begin{equation}
    N_{\rm col} = \int_{t_{\rm form}}^{t_{\rm rec}}dt \,\Gamma_{\rm col}=\int^0_{T_{\rm form}} dT \frac{dt}{dT}\Gamma_{\rm col}\,,
\end{equation}
where $\frac{dt}{dT}$ may be found from the temperature evolution in the early universe \cite{PDGreview},
\begin{equation}
\label{tT}
    (t/{\rm s})(T/{\rm MeV})^2 = 2.4 N^{-1/2}\,.
\end{equation}
Here $N=43/4$ is the effective number of massless degrees of freedom in the QN post formation epoch. Making use of Eqs.~(\ref{GammaCol}) and (\ref{tT}) we estimate the number of collisions of anti-QN with nucleons in the hot primordial plasma
\begin{equation}
\label{Ncol}
    N_{\rm col} = 
    1.1\times 10^{9} B^{2/3} \left(\frac{T_{\rm form}}{41\,\mbox{MeV}}\right)^{3/2}\,.
\end{equation}

Denote by $\kappa<1$ the probability of a nucleon annihilation in the collision with anti QN. Then, the total number of collisions with nucleon annihilation is  $N_{\rm ann}= \kappa N_{\rm col}$.  The survival condition for nucleons and  anti-QN in the early universe
\begin{equation}
    \frac{\Delta B}{B} = \frac{N_{\rm ann} }{B} \ll1
\end{equation}
becomes
\begin{equation}\label{BTconstraint}
     1.1\times 10^{9} B^{-1/3} \kappa \left(\frac{T_{\rm form}}{41\,\mbox{MeV}}\right)^{3/2}\ll1\,.
\end{equation}
It may also be cast in the more convenient form
\begin{equation}
     \frac{B}{\kappa^3} \gg 1.4\times 10^{27}\left(\frac{T_{\rm form}}{41\,\mbox{MeV}}\right)^{9/2}\,.
    \label{Bconstraint}
\end{equation}
This constraint is a refinement of the earlier one $B\gtrsim 10^{24}$ obtained in the work \cite{Survival}, where a number of factors in its derivation have not been taken into account. 

The specific value $T_{\rm form} = 41$ MeV was selected in Ref.~\cite{Survival} to reproduce the measured barion-to-photon ratio $\eta \sim \exp{(-m_p/T_{\rm form}}) \sim 10^{-10}$ which is very sensitive to the value of $T_{\rm form}$. Taking into account that $\eta$ at that time was $(11/4) \cdot 6.2 \cdot 10^{-10}$ we obtain $T_{\rm form} = 46$ MeV.  Independent calculation of  $T_{\rm form}$ in Ref.~\cite{Survival}, based on the consideration of QN dynamic, is not that accurate.

The constraint (\ref{Bconstraint}) involves the unknown coefficient $\kappa<1$ describing probability  of the proton annihilation during collision with anti-QN. Value of $\kappa$, consistent with the limits on $B$ in Eq. (\ref{BTconstraint}) from the early universe data, is 
\begin{equation}
     \kappa \ll 0.8 \cdot 10^{-9} B^{1/3}\,.
    \label{kappaconstraint}
\end{equation}
For $B=10^{24}$ this gives $\kappa \ll 0.1$. The relative probability of the proton annihilation on the anti-QN core can hardly significantly exceed that of antiproton annihilation on nuclei. In the latter process, at the energies about ${\cal E}\sim 40$ MeV, the annihilation cross section is about the factor 0.6 off the total cross section, $\sigma_{\rm ann}= \kappa \sigma_{\rm tot}$, $\kappa\approx0.6$, see, e.g., \cite{Antiproton,Antiproton3}. Adopting this value for the parameter $\kappa$ in Eq.~(\ref{Bconstraint}),  we end up the constraint for the baryon number $B\gg 3\times 10^{26}$, which is about two orders of magnitude stronger than that in Ref.~\cite{Survival}.

We may speculate that the annihilation may be suppressed by the axion domain wall reflecting nucleons and, thus, setting $\kappa \ll 1$. According to the scenario proposed in Ref.~\cite{Zhitnitsky2002}, pressure of the domain wall has formed the color-superconducting  state of antiquarks.
%
One may also argue that the quark wave function of the color superconducting state inside QN  is very different from that in the nucleon \cite{Survival}. This may be a reason of the suppression  of the relative annihilation probability in comparison with the proton-antiproton annihilation  reducing value of $\kappa$ from 0.6 for nuclei to  $\kappa <0.1$ for anti-QN. 

Note that $\kappa$  depends on the QN temperature $T_{\rm QN}$ and kinetic energy of the proton. After the universe cooled down below 1 MeV, the relative velocity of anti-QN and protons is about $0.001c$ and temperature of anti-QN $T_{\rm QN}$ is small in comparison with the electrostatic potential on the quark core surface, so  anti-QN  is surrounded by the positron cloud. Positrons produce friction leading to energy loss and capture of slow protons which do not have sufficient energy to escape \cite{FS21}. Another mechanism of the incident proton energy loss is inelastic collision with the antiquark core. After the capture, annihilation of the protons becomes practically inevitable, $\kappa \approx 1$. The incident proton can avoid the annihilation only if it turns into neutron due to the charge exchange process. However, the cross section of this process is usually smaller than the annihilation cross section, see, e.g., \cite{Antiproton}. This is the reason why we have not included $\kappa$ in our calculations for cold molecular clouds. 


\section{Summary}

In this paper, we continue the study of possible manifestations of quark nuggets as dark matter particles. As is shown in our recent papers \cite{FS21,FS2}, there are three main types of radiation produced in annihilations of anti-QNs with visible matter: thermal radiation from positron cloud, $511$ keV line from positron annihilations and 100-400 MeV $\gamma$-photons produced by decaying $\pi^0$ mesons. We investigate the possibility of observing these three types of radiation in the annihilation of anti-QNs in cold molecular clouds in our galaxy and in the Sun atmosphere. 

As a typical representative of cold molecular clouds, we consider the Taurus cloud as it is one of the nearest to the Earth and, hence, is very well studied. Given the gas density in this cloud, we find the thermal emission from anti-QNs is maximized in the visible and near IR spectrum. We estimated the intensity of this light and demonstrated that it could be detected with modern telescopes such as the Hubble space telescope. Note that this light should be present even if there is no background light from stars. 

We demonstrate also that cold molecular clouds should emit $\gamma$-photons with energies in the range 100-400 MeV. These photons are produced by decaying $\pi^0$ mesons originating from matter-antimatter annihilation. We estimated the flux of such photons, compared it with the one observed by Fermi-LAT \cite{Clouds-gamma} and concluded that  it is plausible that annihilation of anti-QNs in giant molecular clouds produces an observable flux of $\gamma$ photons. Formally, our estimate of the photon flux from anti-QNs for the barion charge  $B <  8 \times 10^{31}$ may exceed the photon flux detected by the Fermi telescope \cite{Clouds-gamma}.

Thus, observation of these two types of radiation (thermal radiation in the visible spectrum and $\gamma$ photons) would be a good confirmation for the quark nugget model of dark matter.

Another possible manifestation of this model could be related to the physics of the Sun. In Refs.~\cite{corona1,corona2}, it was conjectured that annihilation of anti-QNs in the Sun atmosphere can fully explain the heating of the solar corona and resolve its high temperature paradox. We demonstrate that the anti-QN annihilation cross section was strongly overestimated in these works, that leaded the authors to wrong conclusions about possibility of complete annihilation of anti-QNs in the solar corona. We show that the anti-QNs cannot lose any significant fraction of their mass along their trajectories in the solar corona. Qualitatively, it may be understood as follows. Assuming that the anti-QNs survived in the harsh environment of the early universe with hot dense plasma, it seems unlikely that they can completely annihilate in very dilute plasma of the solar corona unless some unknown physics is involved. Thus, the hypothesis of the relation of the quark nugget model to the solar corona temperature paradox is not confirmed.

Although the radiation from anti-QNs in the solar corona is negligible, their annihilation in lower layers of Sun's atmosphere may produce observable effects. We show that the thermal radiation from anti-QNs in chromosphere is mainly produced in the x-ray spectrum. This radiation cannot be observed, as the chromosphere is not transparent for this radiation. However, $\gamma$-ray radiation from $\pi^0$ decay penetrates well the chromosphere and may be observed. We estimated the flux of these $\gamma$-photons and compared it with the one observed by Fermi-LAT \cite{Abdo_2011}. We demonstrated that the estimated flux is comparable with the observed one for $B< 1.3\times 10^{27}$. Thus, it is plausible that $\gamma$ photons from Sun's atmosphere may have the anti-QN annihilation origin.

Last, but not least, we revisited the conditions of surviving of anti-QNs in the early universe. We refined a few factors missing in earlier estimates \cite{Survival}, and showed that the anti-QNs can survive in early universe and be observed today if the baryon charge number is constrained as in Eq.~(\ref{Bconstraint}). This constraint seems  stronger than the earlier proposed one $B\gtrsim 10^{24}$ \cite{Survival}. However, our constraint involves the ratio $B/\kappa^3$, where $\kappa$ is an annihilation suppression coefficient. For annihilation of antiprotons on nuclei  this coefficient is of order 0.6, but it may be smaller for anti-QNs. 

One possible reason for a small value of the suppression coefficient $\kappa$ could be a hypothesis \cite{Zhitnitsky2002} that quarks in the quark core are in the color-superconducting state which has an energy gap with respect to the baryonic state. {A significant difference in the quark nugget and nucleon wave functions may lead to a suppression of the annihilation.} 
Another explanation for a small value of the suppression coefficient $\kappa$  could be attributed to the axion-pion domain wall which was responsible for the formation of quark nuggets according to the scenario proposed in Ref.~\cite{AQNformation}. Domain wall can reflect nucleons and prevent their annihilation. Which of these mechanisms may be realized, is, however, a separate problem to be addressed in the future.

We argue that the domain wall should have an (unknown) mechanism of subsequent decay, because any shell creating a pressure on the quarks would only rise the total energy and may make such object unstable for  decays into baryons, e.g. by emitting (tunneling) of quarks forming nucleons outside. Stability of the quark nugget without the domain wall and any external pressure is possible if the energy of the quarks in the color-superconducting state is lower than the energy of the nuclear matter.

We should also note that the suppression of the nucleon annihilation in the early  universe does not mean suppression of the annihilation inside relatively cold matter. Indeed, slow protons  lose energy due to friction inside positron atmosphere surrounding anti-quark nuggets  and are captured since they do not have sufficient energy to escape. Then the eventual  annihilation is inevitable (unless protons are transformed into neutrons). This makes the anti-QN radiation potentially observable.

\vspace{3mm}
\textit{Acknowledgements} --- We are grateful to M. Murphy and A. Zhitnitsky for useful discussions. We thank Rui-Zhi Yang for clarifying comments about some typos in Ref.~\cite{Clouds-gamma}. This work was supported by the Australian Research Council Grants No. DP190100974 and DP200100150 and the Gutenberg Fellowship.


\appendix
\section{Stopping power in solar corona plasma}
\label{AppA}

In Sect.~\ref{sec-corona}, we considered the motion of anti-QNs through the solar corona plasma ignoring the collisional friction of anti-QNs in this medium. In this section, we estimate the stopping power of anti-QNs in the solar corona plasma and compare it with the gravitational attraction force. We also demonstrate that the stopping power of anti-QNs was strongly overestimated in Ref.~\cite{corona2}. 

There are three main effects in plasma which contribute to the stopping power: 1. Proton inelastic scattering with capture and subsequent annihilation;
2. The elastic proton scattering off the Coulomb potential of the anti-QN and 3. The collective plasma effects caused by a moving charged particle. We estimate these effects separately.

\subsection{Proton inelastic scattering}

In Sect.~\ref{sec-corona-total}, we demonstrated that the capture cross-section of protons scattering off an anti-QN is $\sigma_{\rm col} = \gamma \pi R^2$, where $\gamma=1.1$ is the enhancement factor (\ref{gamma-factor}). Along the path $\Delta z = v\Delta t$, the captured protons reduce the anti-QN momentum as
\begin{equation}
    \Delta p = m_p v n \sigma_{\rm col} \Delta z
    = m_p v^2 n \sigma_{\rm col} \Delta t\,.
\end{equation}
The corresponding contribution to the stopping power is 
\begin{equation}
    -\frac{d{\cal E}_{\rm cap}}{dz} = m_p v^2 n \sigma_{\rm col}\,.
\end{equation}
For a typical anti-QN velocity $v = 2\times 10^{-3}c$ and maximum solar corona particle density $n=10^{10}\mbox{ cm}^{-3}$, we estimate
\begin{equation}
    -\frac{d{\cal E}_{\rm cap}}{dz} = 1300 \mbox{ keV/m}\,.
    \label{Fcap}
\end{equation}

\subsection{Proton elastic scattering}

Consider a proton scattering elastically off the anti-QN with the charge $Q=Ze$. Assume that in the anti-QN rest frame the proton moves with the velocity $v=2\times 10^{-3}c$. We consider classical scattering off the Coulomb potential $U = Ze/r$ which corresponds to the force ${\bf F} = Ze^2{\bf r}/r^3$. Let $F_\parallel$ and $F_\perp$ be components of the Coulomb force parallel and orthogonal to the velocity vector of the incident proton. It is possible to show that $F_\perp$ gives the main contribution to the stopping power while $F_\parallel$ averages out and may be neglected. The force $F_\perp$ is responsible for the following momentum transfer
\begin{equation}
\begin{aligned}
    \Delta p_\perp &= \int F_\perp dt = \int F_\perp \frac{dz}{v} \\
    &=\int_{-\infty}^{\infty} \frac{Ze^2}{z^2+b^2} \frac{b}{\sqrt{z^2+b^2}}\frac{dz}{v}
    =\frac{2Ze^2}{bv}\,,
\end{aligned}
\end{equation}
where $b$ is the impact parameter. The corresponding energy transfer in the scattering of one proton is
\begin{equation}
    -\Delta{\cal E}_1(b) = \frac{\Delta p_\perp^2}{2m_p} = \frac{2Z^2 e^4}{m_p b^2 v^2}\,.
\end{equation}

Now consider a number of protons in the cylindrical layer $d N_p = n 2\pi b db dz$. For all these particles, the energy transfer is 
\begin{equation}
    -d{\cal E}_{\rm scatt}(b) = -\Delta{\cal E}_1 (b) dN_p = \frac{4\pi n Z^2 e^4}{m_p v^2} \frac{db}{b}dz\,.
\end{equation}
Integrating over the impact parameter from $b_{\rm min}$ to $b_{\rm max}$, we find the stopping power
\begin{equation}
    -\frac{d{\cal E}_{\rm scatt}}{dz} = \frac{4\pi n Z^2 e^4}{m_p v^2} \ln\frac{b_{\rm max}}{b_{\rm min}}\,.
\end{equation}
Here $b_{\rm min}$ may be identified with $b_{\rm min} =\sqrt{\gamma} R \approx 10^{-5}$ cm, because $\gamma \pi R^2$ is the proton capture and annihilation cross section. The upper limit may be identified with the Debye length, $b_{\rm max} = \lambda_D \approx 0.07$ cm (see Eq.~(\ref{lambdaD}) below). 

The anti-QN electric charge number $Z$ may be estimated from the condition that the potential energy of bound electron orbiting anti-QN is twice the ionization potential, $|U(R)| = \frac{Ze^2}{R} = 2k_B T$ $\Rightarrow$ $Z = 14000$ for $T\simeq 100$ eV. For these parameters, we find the stopping power
\begin{equation}
    -\frac{d{\cal E}_{\rm scatt}}{dz} = 120 \mbox{ keV/m}\,.
    \label{Fscatt}
\end{equation}

\subsection{Collective plasma effects}

The solar corona plasma has particle number density $n = 10^{10}\mbox{ cm}^{-3}$ and temperature of order $k_B T = 100$ eV. The corresponding plasma frequency and the Dedye screening length are:
\begin{eqnarray}
    \omega_{\rm pl} &=& \sqrt{\frac{4\pi n e^2}{m_e}} = 3.7 \times 10^{-6}\mbox{ eV} = 5.6 \mbox{ GHz}\,,\\
    \lambda_D &=& \sqrt{\frac{k_B T}{4\pi n e^2}} = 0.07 \mbox{ cm}\,.
    \label{lambdaD}
\end{eqnarray}
The typical electron velocity in this plasma is $\langle v_e^2 \rangle^{1/2} = \sqrt{3k_B T/m_e} = 0.02 c$. Thus, the incident anti-QN particle is slow as compared with the electron velocities, $v\ll \langle v_e^2 \rangle^{1/2}$ for $v = 2\times 10^{-3}c$. In this regime, the collective effects in plasma make the following contribution to the stopping power (see Ref.~\cite{plasma}):
\begin{equation}
\label{stopping}
-\frac{d {\cal E}_{\rm pl}}{dz} = \frac{Z^2 e^2 v \omega_{\rm pl}^2}{3\sqrt{2\pi}} \left( \frac{m_e}{k_B T} \right)^{3/2}
\ln \left( \frac{k_B T n^{2/3}}{3.17 m_e \omega_{\rm pl}^2} \right) \,,
\end{equation}
with $Z=14000$ being the charge number of anti-QN in plasma. For this anti-QN charge, we find the stopping power
\begin{equation} 
    -\frac{d {\cal E}_{\rm pl}}{dz} =86\mbox{ keV/m}\,.
    \label{Fpl}
\end{equation}

Summing up (\ref{Fcap}), (\ref{Fscatt}) and (\ref{Fpl}), we find the total stopping power of anti-QN moving through the solar corona plasma
\begin{equation}
    F_{\rm stop} = -\frac{d {\cal E}_{\rm cap}}{dz} 
    -\frac{d {\cal E}_{\rm scatt}}{dz}
    -\frac{d {\cal E}_{\rm pl}}{dz} = 1.5 \mbox{ GeV/km}\,.
    \label{Fstop}
\end{equation}
Given that the anti-QN kinetic energy is of order $2\times 10^{18}$ GeV, the stopping power (\ref{Fstop}) is unable to change significantly the anti-QN velocity along its path in the solar corona.

\subsection{Comparison with gravitational attraction}

The force of gravitational attraction of an anti-QN near the Sun surface is estimated as
\begin{equation}
    F_{\rm grav} = \frac{G m_{\rm QN} M_\odot}{R_\odot^2}
    =3\times 10^{12}\mbox{ GeV/km}\,.
    \label{Fgrav}
\end{equation}
Thus, the stopping power of anti-QN in the solar corona (\ref{Fstop}) is negligible as compared with the gravitational attraction to the Sun.

\subsection{Comparison with the results of Ref.~\cite{corona2}}

In Ref.~\cite{corona2} it was argued that the anti-QN ionization may be much stronger, $Z\sim 10^8$. For this charge, Eq.~(\ref{stopping}) yields much greater value of the stopping power $-\frac{d{\cal E}}{dz} =4500$ GeV/m. This value is, however, still lower than overestimated stopping power in Ref.~\cite{corona2}, $-\frac{d{\cal E}}{dz} = \frac\pi2 R_{\rm eff}^2 m_p n v^2 \approx 6\times 10^{7}$ GeV/km, calculated with $R_{\rm eff} = 0.1$ cm. Thus, the stopping power for an anti-QN moving in the solar corona was strongly overestimated as compared with our result (\ref{Fstop}). However, even this overestimated result is negligible as compared with the gravitational attraction force (\ref{Fgrav}).


%

\end{document}